\documentclass[lzlncs,11pt]{article}

\usepackage{amsmath}
\numberwithin{equation}{section}

\usepackage[utf8]{inputenc}
\usepackage[english]{babel}
\usepackage{graphicx}
\usepackage{xcolor}
\usepackage{float}

\usepackage{ifthen}
\newboolean{showNotes}
\setboolean{showNotes}{false}

\DeclareFontFamily{U}{mathx}{\hyphenchar\font45}
\DeclareFontShape{U}{mathx}{m}{n}{<-> mathx10}{}
\DeclareSymbolFont{mathx}{U}{mathx}{m}{n}
\DeclareMathAccent{\widebar}{0}{mathx}{"73}

\usepackage{amssymb,amsmath}

\usepackage{newtxtext}

\usepackage[T1]{fontenc}

\usepackage[normalem]{ulem}
\usepackage[color=yellow!20]{todonotes}
\usepackage{soul}
\usepackage[linktoc=all]{hyperref}
\usepackage{paralist}
\usepackage{mathrsfs}
\usepackage{listings}
\usepackage{lscape} 
\usepackage{dashbox} 

\usepackage{dirtytalk} 
\usepackage{booktabs}
\usepackage{pifont}
\usepackage{enumitem} 

\usepackage[margin=1in,a4paper]{geometry}
\usepackage[framemethod=TikZ,innerleftmargin=5pt,innerrightmargin=5pt]{mdframed}
\usepackage[listings,skins,breakable]{tcolorbox}
\usepackage{bm}
\usepackage[mathscr]{euscript}
\usepackage[all]{nowidow}

\usepackage{sectsty}
\allsectionsfont{\normalfont\sffamily\bfseries}

\usepackage[T1]{fontenc}

\newtheorem{definition}{Definition}
\usepackage{biblatex}

\title{Mempool Privacy: An Economic Perspective}

\author{Antoine Rondelet and Quintus Kilbourn}
\date{\today}

\bibliography{references}

\begin{document}

            
          
\maketitle
    \begin{abstract}
          A growing body of literature is aimed at designing private mempools in blockchains. The ultimate goal of this research is addressing several phenomena broadly classed under MEV with sandwich attacks as the canonical example. The literature has primarily viewed MEV as a problem arising from oversights in distributed systems and cryptographic protocol design and has attempted to address it with the standard tool sets from those disciplines. This paper argues that the impact of private mempools on markets and agent incentives renders analyses that do not consider the economic lens incomplete. The paper presents several observations across blockchains and traditional finance to justify this argument and highlight specific dynamics for future study.
          
        \end{abstract}
\section{Introduction}
Privacy has been a theme intertwined with blockchains since their inception in the form of Bitcoin \cite{bitcoinWhitepaper}. Blockchains have both been deeply connected with financial applications - use cases which often necessitate privacy - and relied heavily on cryptographic tools and the cryptographic community. One particular area of development in this larger push has been ensuring the privacy of transactions which have not yet been executed, dubbed "mempool privacy."

While mempool privacy is naturally relevant when user actions are intended to be hidden indefinitely, much recent work has focused on an ephemeral notion of privacy which sees information about transactions partly or fully revealed after being executed. The main setting for such privacy is smart contract blockchains of which Ethereum is emblematic. 

The importance of mempool privacy to smart contract blockchains can be understood from two perspectives: the functioning of blockchains as an "infrastructure layer" and the optimal functioning of the financial markets in the "application layer" above. The importance of mempool privacy to smart contract blockchain infrastructure became clear in 2019 with the publication of "Flash Boys 2.0" \cite{flashboys}, which introduced the concept of MEV (Maximal Extractable Value). The paper detailed how information about unexecuted transactions in the Ethereum mempool, in part, was leading to a variety of behaviours which congested Ethereum, increased the incentive for chain forks and saw some users profiting at the expense of others. These and similar phenomena, some of which lead to the crash of other blockchains \cite{solanaDoS,avaxMEV,avaxSnowman}, propelled the study and mitigation of MEV into a field in its own right \cite{fanMEV} - a field which has naturally placed an emphasis on mempool privacy. 

To give context on scale, we can turn to transfers of value attributed to MEV. In the last year, over 300 000 ETH (just under half a billion dollars according to the prices at writing, Sept '23) has been transferred to the Ethereum validator set due to MEV activity \cite{flashbots-dashboard}, a lower bound of the scale of total value transfer.

From a traditional financial perspective, the importance and impact of privacy of unexecuted financial operations is a well-studied topic in practice, in the literature and the regulatory frameworks. For example, many large exchanges offer trade types like iceberg orders \cite{icebergs} which allow traders to obscure the quantity while exposing the price at which they are willing to trade. Other trading venues, known as dark pools, differentiate themselves through privacy \cite{darkpools}. Financial markets researchers have also developed a rich literature investigating the relationship between privacy and market outcomes. For example, some research has investigated the market properties under which foregoing trade privacy may be desirable \cite{sunshine,sunshine2,sunshine3,sunshine-trading} while other research has investigated the optimal means by which a competing trader's lack of privacy can be exploited \cite{arbingarbers,predatoryTrading}. On the regulatory front, regulations such as FINRA 5270 in the United States strictly prohibit financial service providers from acting on information about unexecuted trades. 

 Examples from the traditional financial industry also provide insight into the extremes to which traders will go to acquire access to information which has not been incorporated into the market. For instance, in the recent past, some firms have paid to acquire satellite images of Walmart parking lots in order to count parked cars as an indication of stock value \cite{walmart} while hundreds of millions of dollars were spent laying a cable between New York and Chicago for a 3ms reduction in roundtrip time between exchanges \cite{frequentBatchAuctions}. Not only should blockchain designers expect similar dedication to information extraction, we should expect small details to have substantial impact on the market structure.
 

 Just like MEV in the blockchain context, the lessons from traditional finance point to the pivotal role which the privacy of unexecuted transactions can play in decentralised finance. In some sense, the economic lens has been applied to what is now being labelled as "mempool privacy" for decades. However, in another sense, the economics of mempool privacy have gone with little attention. Traditional financial settings do not have validators who are able to produce empty blocks, gossip networks, decryption phases or near-Turing-complete transactions. Instead of traditional finance's regulated third parties, blockchain designers have goals of censorship resistance and decentralisation. The lack of distributed systems and cryptographic context in this relevant economic research and the overemphasis on distributed systems and cryptographic design in blockchain-native work points to a gap in understanding which must be filled.

\subsection{Our Contributions}
Thus far, formal research into mempool privacy has mostly taken a distributed systems or cryptographic lens. As the issues being addressed are fundamentally economic in nature, solutions that do not consider the economic perspective may fail to address or aggravate targeted issues. This paper relates and draws from observations in blockchains in production as well as financial markets to highlight several economic dynamics which require further study and consideration when designing private mempools. In particular, the paper addresses (1) the impact of mempool privacy on user behaviour and how this impacts network congestion and security guarantees; (2) the impact of mempool privacy on validator incentives as they pertain to censorship, decryption delays and collusion and; (3) the impact of mempool privacy on DeFi markets, using liquidations as a motivating example.
\section{Preliminaries}
\subsection{Mempool Privacy}
This paper deals with private mempools and the associated concept of "mempool privacy" or "pre-execution privacy". Unlike in traditional finance, in which execution and settlement have exact legal meanings \cite{yale-reg,settlement-finality}, the terms are harder to pin down in the blockchain context. As such, we work with an operative definition of execution.
\begin{definition}
    A blockchain transaction is classified as \textbf{executed} if the transaction is valid and no single party or small group of parties is able to change the outcome of the transaction, where the outcome of the transaction is a transition in the blockchain state machine. 
\end{definition}
A transaction is thus "pre-execution" if the transaction is valid, but a validator or other party is able to influence which state transition it engenders in the view of the broader network. There certainly are some important unanswered questions surrounding notions of execution, but the definition is sufficient for the purposes of this paper. 

Having defined execution, we are now able to define mempool privacy.
\begin{definition}
    A blockchain satisfies the property of mempool privacy if the sender of a transaction is able to hide some or all of the content of a transaction until it is considered executed. 
\end{definition}

Clearly, this is not a rigorous cryptographic definition and leaves the exact nature of the "content" being obscured under-defined. This vagueness is partly intentional as different mempool designs target slightly different properties and the given definition is sufficient for the purposes of this paper. A useful framing for the notion of mempool privacy is as relatively more private compared to the status quo on the most widely used blockchains like Ethereum. The status quo being that transactions are propagated completely in plaintext with the state transitions they engender easily accessible to the block producer (i.e. the validator or miner selected who produces the next block) if not a larger set of agents. A similar definition of mempool privacy can be found in \cite{DBLP:journals/iacr/BebelO22} which specifically requires that transactions are encrypted and emphasises inclusion in a block as opposed to execution.

\subsection{Motivations For Mempool Privacy}
Although not exhaustive, there are two commonly-cited reasons that mempool privacy is desirable, all of which are often lumped under the umbrella of MEV.

\subsubsection{Asymmetries}\label{ssec: asymmetriesProb}
Currently, popular blockchains use consensus algorithms which allocate the role of producing the next block to a single validator (or miner in proof of work algorithms). This is true of Gasper on Ethereum\cite{gasper}, Tendermint or CometBFT in Cosmos chains \cite{tendermint}, Solana consensus\cite{solanaWhitepaper}, Hotstuff\cite{hotstuff}, Avalanche consensus and many more. The absence of mempool privacy in combination with an ephemeral monopoly over transactions ordering and inclusion creates an information and power asymmetry between the block producing node and users who necessarily must expose the content of their transactions to the block producer before they are executed. This allows the block producer to engage in "frontrunning" and "sandwiching" \cite{cfmm_sandwiches} among other activities.

Proponents of mempool designs which seek to curb this asymmetry cite unfairness as a primary motivation \cite{themis,aequitas,dagMev}. Such arguments have given rise to notions of fairness such as \\

\indent \textbf{Receive-Order Fairness} which considers ordering transactions in the order by which they are received by the network as fair \cite{aequitas}.\\
\indent \textbf{Blind-Order Fairness} which considers transaction ordering being conducted without any knowledge of the content of transactions as fair \cite{dagMev}. Note, that this is less restrictive than receive-order fairness as the block producer is still able to reorder transactions.\\

Fairness-based criterion are not unique to blockchain markets and are often discussed in traditional markets, often with regard to the protection of unsophisticated "retail" traders \cite{sec-fairness,cnbc-retail-stock}.

The aforementioned information asymmetry can also be deemed undesirable from an economic efficiency or market neutrality perspective. As an example of economic inefficiency, sandwich attacks can increase the cost to trade, inhibiting efficient allocations. Censorship, broadly defined as the delayed or denied execution of a valid transaction, is a commonly presented example of market neutrality being violated. Such censorship could be politically motivated \cite{wahrstätter2023blockchain} or driven by economic reward \cite{censorshipAuctions}. Delaying posting of collateral, liquidations \cite{liquidations} or oracle updates are basic examples of behaviours which can both be profitable for the validator and threaten the correct functioning of market deployed on the blockchain. 

\subsubsection{Externalities of Strategic Use}\label{ssec:PGAs}
Current blockchain implementations are resource constrained. As observed in \cite{flashboys}, public mempools can lead to dynamics such as Priority Gas Auctions (PGAs) in which users pursuing the same profitable opportunity on the blockchain (e.g. a liquidation) will rapidly respond to each other's pending transactions in an attempt to outbid competition pursuing the same opportunity in what resembles an English auction. The net effect is a heavily congested network layer, blocks filled with many failed transactions and elevated fees for accessing the blockchain. 

\subsection{Encrypted Mempool Design Space}
It is useful to highlight several important dimensions of the private mempool design space. Naturally, there are many details to consider, but the dimensions listed below are relevant insofar as they relate to economic considerations outlined in \ref{sec:economics}. \\

\subsubsection{Underlying Cryptography}
There are several cryptographic primitives which have been utilised in different design proposals. Outside of the receive-ordering work, all encrypted mempool designs that the authors are aware of attempt to ensure decryption only occurs after the content of a block has been finalised (i.e. after the transaction has been executed). As elaborated in \ref{ssec: litreview}, designs in which users encrypt to a threshold public key \cite{DBLP:journals/iacr/BebelO22,dagMev,helix,threshold-anoma,threshold-penumbra} are popular as threshold honesty assumptions often resemble the honesty assumptions of the underlying consensus protocol. Trusted Execution Environments (TEEs) have also been proposed in conjunction with commit reveal schemes \cite{PROF,SUAVE}. The use of timelock encryption \cite{timelock-frontrunning} and commit-reveal schemes \cite{flashfreezing,breakingTheChains}  have also been proposed.

Importantly, these primitives can be combined and used in different ways. An in-depth comparison of this component of the design space falls outside the scope of this paper as many of the arguments raised in \ref{sec:economics}, the core of this paper, apply independent of underlying cryptographic primitives and schemes.\\

\subsubsection{Denial-of-Service Resistance}\label{ssec: DoS}
 As blockchains generally allow any actor to enter messages into the mempool, nodes which propagate these messages are susceptible to being overloaded with very high message loads. Cheap deduplication and validity checks are common techniques to defend against such forms of attacks~\cite{harden-network-sec}. While the exact nature of the validity check varies between blockchain implementations, the general approach of verifying that a transaction pays a requisite fee is universal as this compensates the network for resource usage and imposes costs on would-be attackers.

Unfortunately, a naive implementation of mempool privacy in which too much information is obscured prohibits such validity checks. An easy workaround is to make sender account information available so that ownership of sufficient funds to pay fees can be efficiently verified. This comes at the expense of privacy for the sender's account, e.g. \cite{DBLP:journals/iacr/BebelO22}.
Another approach which avoids this tradeoff is to attach a cryptographic proof that a ciphertext corresponds to a valid transaction \cite{zcash-specs}. However, computing and verifying such proofs may introduce overhead with significant impact in latency-sensitive trading environments. Additionally, special restrictions must be put in place to ensure that the same funds are not used to generate multiple proofs exceeding the value of the funds in aggregate. We have yet to see a proof-based DoS-resistance mechanism that is suitable for smart contract blockchains. 

 Considering means for DoS resistance points to a larger question of what information to leave accessible and what to obfuscate. Tradeoffs around this are discussed further in \ref{sec:economics}.

 \subsubsection{Optionality}
 While the concept of shielded and transparent transactions exist in certain private blockchains such as ZCash \cite{zcash-specs}, the idea of allowing users to choose the privacy parameters of their transactions has not been thoroughly discussed in the context of smart contract chains. Standard motivations for this kind of optionality are based around users seeking to comply with regulatory requirements, but there are additional motivations in terms of users avoiding costly overhead from encryption schemes, seeking faster settlement and intentionally propagating information to a market of counterparties and competitors as discussed below. A specific setting in which users are given a choice to opt in or out of a private mempool is studied in \cite{litToDark}.

\subsection{Private Mempool Literature}\label{ssec: litreview}
Several different designs for private mempools for smart contract blockchains have been proposed. Among these, Ferveo \cite{DBLP:journals/iacr/BebelO22} proposes a threshold encrypted mempool which aligns closely with the security assumptions of the Tendermint consensus protocol\cite{tendermint}. A similar paper shows how threshold encryption can be used to provide mempool privacy in DAG-based blockchains \cite{dagMev}. Addition of threshold encrypted mempool to production blockchains has also been proposed \cite{shutterisedbeaconchain,threshold-penumbra}.

A separate class of solutions have proposed the use of TEE's. Some solutions advocate for the use of TEE's external to the validator set in combination with a commit-reveal scheme \cite{PROF,SUAVE}, while some blockchains which are already live require validators to run TEEs themselves \cite{oasis,secretNetwork,obscuro}. Other work has proposed timelock encryption \cite{timelock-frontrunning} while commit-reveal schemes are proposed in \cite{flashfreezing} and in \cite{breakingTheChains} which notably operates under a rational model. 

A related branch of literature pursues receive-order fairness, which would also give a notion of mempool privacy \cite{aequitas,themis,order-fair-ic3}. Notably, the feasibility and welfare implications of such a property have been met with criticism \cite{fifoWelfare,spectraMEV}. 

\section{Economic Considerations}\label{sec:economics}
This is the main section of the paper which outlines some economic dynamics around private mempools which are both not well understood and impactful, necessitating further research. While specific implementations of blockchains with private mempools will have their own idiosyncracies, the questions and observations below are intended to apply broadly and should serve as inspiration for research tailored to particular settings.

\subsection{User Incentive Compatibility}
The implementation of an encrypted mempool does not imply that users will engage with the protocol as intended. Sending unencrypted information to middlemen, generating large numbers of transactions with a high probability of failure on execution or exposing information about a ciphertext to the block producer or the public are all examples of such behaviour. 

\subsubsection{Inefficient Resource Consumption}\label{sssec: specTX}
In \ref{ssec:PGAs} PGA's are listed as a phenomenon partly caused by the public nature of Ethereum's mempool, which strains limited network resources. In priority gas auctions, rapid submission of transactions in response to those submitted by competitors can congest the gossip network. The obfuscation of competing transactions can address this by removing the information which triggers competitive responses, although, as explained in \ref{ssec: DoS}, some proposed approaches to DoS-resistance do not obscure key information like fees, which can trigger this kind of iterative bidding behaviour.

That said, addressing the escalating bidding of PGA's is but one part of the larger problem of competitive dynamics wasting network resources. PGA's see users speculatively submitting transactions knowing that there is some probability of their transaction capturing a profitable opportunity. As \cite{flashboys} notes, the consequence of this speculation is that blocks of limited size can contain significant numbers of failed transactions. The uncertainty as to whether a user will be able to exercise an opportunity in PGA's depends on bidding behaviour and network latency of competitors. When privacy is introduced, there is additional uncertainty about the existence and nature of opportunities in the first place, potentially leading to more failed transactions being included in blocks (which also add to gossip network congestion). 

To find concrete examples of this phenomenon, we can turn to Solana \cite{solanaWhitepaper} a low-latency blockchain which does not have a public mempool as transactions are directly forwarded to the block producer. Instead of waiting for a transaction which engenders an arbitrage or liquidation opportunity to be executed and only then reacting, sophisticated users of Solana optimistically send transactions which attempt to capture an opportunity before its existence is even publicly known \cite{solanaSpam}. This is part of a larger trend, partly fueled by low fees, in which 95\% of Solana arbitrage transactions fail, making up more than 20\% of transactions on Solana in 2022 according to a Solana industry leader\cite{jito_failed}.

Further research is needed to understand the relationship between mempool privacy, fee markets and speculative transaction submission. The most notable responses from industry have come in the form of the block builder role in Ethereum which commonly provides the service of simulating and excluding failed transactions and Jito Labs' Block Engine on Solana which provides a similar service and artificially creates a public mempool where there was none before \cite{jito}.

\subsubsection{Intentional Information Revelation and Intermediaries}\label{intentionalRevelation}
Phenomena like sandwiching \cite{cfmm_sandwiches} have been taken to motivate a position that privacy must necessarily improve user execution. However, upon further consideration this does not hold in general. Consider the basic example of four trades, two buys and two sells of the same asset pair, being ordered in a block. Blind ordering attaches some probability to the ordering of <buy, buy, sell, sell> which may see the second and fourth transaction fail as prices have been moved out of acceptable ranges. An alternating ordering doesn't have this effect, leaving users an incentive to circumvent blind ordering. 

This example should be enough to illustrate the basic point that users may prefer less-than-full privacy, but there is actually a wealth of evidence to support this point. Traditional finance is full of examples of pre-execution information revelation. Some traders may be incentivised to publicly announce their trade intentions in what is known as "sunshine trading" \cite{sunshine, sunshine2,sunshine3} while request-for-quote (RFQ) systems often involve traders exposing some information like asset pair and quanitity while obscuring the direction of their trade as part of a counterparty discovery process. 

As trading environments are replicated in the blockchain setting, we should expect these incentives to reveal pre-execution information to be replicated as well. The question is how this would occur concretely.

 Structures like the RFQ's mentioned above are already well established in blockchain. In fact, very large amounts of DeFi volume on Ethereum flow through at least one intermediary that provides some services to improve execution quality, often never entering the public mempool \cite{smgBuilders,builderProfiles,blocknativePrivate}. Block builders like Flashbots or Titan builder; and auction systems like 0x, 1inch, CoWswap, MEV-share and MEV-blocker are prevalent examples of such services. Notably, Uniswap, the largest decentralised exchange on Ethereum, recently announced the deployment of such an RFQ system to service all users interacting with their front end \cite{uniswapX}.
 
On one hand, the existence and usage of such intermediaries is beneficial to blockchain systems because of the additional features and improved execution quality presented to users. On the other, however, the usage of these intermediaries can erode the guarantees blockchains aim to provide users. The properties of decentralisation may be weakened if effective use of the applications deployed on the blockchain rely on the services of a trusted intermediary, forcing users to choose between execution quality and security. This isn't an effect isolated to individual users as there are network effects to trading activity, with popular venues being more appealing \cite{liquidityNetworkEffect}.

\textbf{The connection between these intermediaries and mempool privacy is that the degree and nature of reliance on intermediaries depends on the structure of the underlying protocol.} Intermediaries are unlikely to replicate functionality the underlying blockchain already provides. For instance, users who currently rely on intermediaries for forms of trusted privacy can use the public mempool under mempool privacy. Altneratively, users who seek to credibly reveal information about their unexecuted transactions are more likely to use intermediaries to do so if the underlying protocol doesn't support the (full or partial) revelation the user is after.

Similarly, the trust assumptions implied when using an intermediary will vary depending on details of the underlying blockchain. It is common for intermediaries to aggregate transactions (or cryptographic messages of other formats) from users into what is known as "bundles" on Ethereum before passing these on to other intermediaries (e.g. \cite{erc4337}). Currently, on Ethereum, the second intermediary (or an attacker that compromises the intermediary \cite{unbundlingAttack}) is in a position to decompose bundles, using these for unintended state transitions. If the underlying blockchain's signature scheme supports a form of aggregation, users are protected from unbundling.

Such a relationship between the protocol and surrounding intermediaries calls for protocol design which explicitly reasons about the existence of intermediaries including users' incentive to rely on them and the properties that the entire system can guarantee. 
A clear starting point in varying private mempool designs would be providing users meaningful degrees of freedom in choosing what information is obfuscated, perhaps allowing for ciphertext transactions to be accompanied with proofs about properties of the preimage. This allows both for control over information revelation to the market at large, but also to intermediaries.

Some work on this intermediary-aware blockchain interface has already begun in industry. The Ethereum community has started working on standards to facilitate the easy use of some intermediary services in the form of proposals like ERC4337 (a message standard that uses BLS signatures) and proposer-builder separation (a commit-reveal scheme) \cite{erc4337,4337mempool,ePBS}. The Anoma architecture has also been designed to reason explicitly about private pre-execution counterparty discovery through the use of tools like zero-knowledge proofs \cite{anomaWP}. 

\subsection{Validator Incentive Compatibility}
Validators (or miners) perform essential tasks such as acting as block producers, voting on blocks and, under some designs, acting as a threshold decryption committee. Apart from \cite{litToDark} which investigates the setting in which validator participation in private mempools is optional, little work has gone into understanding how introducing mempool privacy impacts validator incentives. 

An important delineation to make is that mempool privacy does not necessarily prevent block producers from influencing the inclusion and ordering of transactions and being compensated for doing so. For example, a user may prefer to have their transaction execute first in a block as this guarantees access to identified opportunities. Ordering policies are outside the scope of this work.

\subsubsection{Censorship}
In \ref{ssec: asymmetriesProb}, the connection between neutrality, in particular the inclusion of valid transactions, and mempool privacy was outlined. Apart from the obvious point that it is more difficult to censor transactions according to certain criteria if the information for evaluating those criteria is not available, there are two dynamics to highlight. 

The first thing to point out is that not all information is necessarily obscured. This could be due to the protocol design (e.g. for DoS resistance), but also due to users voluntarily exposing information in pursuit of better execution guarantees (see \ref{intentionalRevelation}). The latter may force users who are targets of censorship to resort to less efficient (fully private) means of using the blockchain, an improvement over censorship but still short of a neutral platform.

The second notable dynamic is that block producers may be incentivised to censor broad groups of transactions if censoring individual transactions is not feasible. Block producers are generally afforded the option to produce an empty block, which represents an extreme option. Other forms of information available can result in finer selection, which in turn changes users' incentives to participate in privacy schemes as they seek to reveal information that differentiates their transactions from those targeted by a censorship policy. 

Research is needed to understand this dynamic and how it interfaces with other aspects of the blockchain design, such as the network behaviour when an empty block is produced (e.g. is the leader rotated? is the next leader known?) or where transactions fees go (e.g. Ethereum's EIP1559 mechanism does not pay the block producer, reducing incentives for transaction inclusion).

\subsubsection{Collusion}
Numerous designs across academia and industry \cite{threshold-anoma,threshold-penumbra,threshold-shutter,threshold-sikka,DBLP:journals/iacr/AsayagCGLRTY18,DBLP:journals/iacr/BebelO22,dagMev} have proposed the use of threshold encryption schemes to implement mempool privacy. The justification is often that honest-majority thresholds can be set to mimic those of the underlying protocol (e.g. $\frac{2}{3}$ for Tendermint). This justification, however, does not sufficiently weigh the fact that some deviations from honest behaviour such as double-signing, are detectable and attributable and can be disincentivised through the threat of seizing collateral ("slashing") and reputational damage; while other deviations such as early decryption of encrypted transactions can be neither, removing such disincentives. 

One research direction is modeling these external incentives and proposing methods for monitoring for misbehaviour, as has been proposed in industry for other forms of misbehaviour \cite{skipAccountability}. 
Traditional financial markets have many examples of collusion to manipulate markets for profit.
Examples include the Libor scandal \cite{libor} foreign exchange rate manipulation scandal \cite{forexCollusion}, municipal bond rigging \cite{municipal-bond-bid-rigging}, ISDAFIX scandal \cite{isdafix,isdafix-cftc} and metals price-fixing scandal \cite{doj-jpm-precious-metal,precious-metals-collusion}.

\subsubsection{Information Asymmetries}
Due to non-zero propagation delays, some party must learn about the decrypted information first. Due to the valuable nature of information in financial markets, this may constitute a valuable asymmetry. Thus, there is a clear incentive for parties who learn information first to delay propagation, increasing the length of this asymmetry. A similar dynamic has already been studied in the context of Ethereum in which a block producer is incentivised to delay the production of their block as this increases its value due to the continued arrival of transactions \cite{EthTimingGames}. The conclusion was that the incentives for this behaviour were present, although not currently acted upon. 

Extensions of this research are required to evaluate various private mempool designs, the conditions under which there are incentives to delay information revelation and the means to disincentivise such delays.

\subsection{Market Impact}
The interactions between mempools and the applications which ultimately rely on them certainly depends heavily on the nature of the applications which are deployed. A large body of financial literature investigates the impact of the introduction of private trading venues, "dark pools", on price discovery, trader welfare and price spreads (difference between best buy and sell prices) \cite{darkPoolPricing,darkpools,darkPoolWelfare}. The extent to which this research translates well to the blockchain setting depends on model assumptions and blockchain market design. There is, however, good reason to believe that the introduction of privacy will impact decentralised exchanges. Other financial primitives like lending are certainly also likely to be affected, as discussed below. 

\subsubsection{Liquidations}
Lending is a fundamental financial activity which is common in DeFi in the form of applications like Aave and Compound \cite{aave,compound}. These applications rely heavily on the use of collateral to secure loans. As the value of the collateral relative to the loaned assets fluctuates with time, it is often necessary for these applications to sell collateral to clear bad debt in what is known as a \emph{liquidation} \cite{liquidations}, a process which relies on external agents to purchase the collateral. In times of rapid price movements in which the relative value of the collateral is steeply declining, it is important for lenders to be able to liquidate collateral quickly in order to avoid bad debt. 

In blockchains with transparent mempools like Ethereum, it is common for liquidations to be executed in the same block in which the collateral becomes eligible for liquidation. Usually, this is due to an oracle update or a price-moving transaction. However, in settings where the existence of transactions triggering price movements are only known after they have been executed, liquidators can only act in the next block after discovering this information while the price on other venues like centralised exchanges continue to move. 

Naturally, there are some counter-forces to these effects such as the speculative transactions explained in \ref{sssec: specTX}. The bottom line, however, is that less efficient liquidations necessitate higher collateral requirements reducing capital efficiency (i.e. the utility of the lending protocol) and, since mempool privacy can impact the efficiency of liquidations, the relationship between liquidations and privacy requires further investigation.



\section{Conclusion}
This paper has argued that private mempools must be analysed through an economic lens. Such an economic analysis has been motivated by highlighting how the introduction of mempool privacy can impact the incentives and behaviour of agents that use and comprise a blockchain in undesirable ways, threatening system security and the efficiency of DeFi markets. The paper has aimed to stimulate research at the intersection of cryptography, distributed systems \textbf{and} economics by proposing several open areas of study in the design of private mempools.


\newpage
\printbibliography

\appendix

\section{A Case Study: Osmosis}

\subsection{Osmosis Overview}

Osmosis is an application-specific automated market maker blockchain specialized for decentralized exchange transactions rather than arbitrary state transitions, built on the Cosmos ecosystem \cite{cosmos-wp,consensys-osmosis}. The Osmosis chain is also home to Mars Protocol \cite{MarsProtocol}, a collateralized lending platform.

Over time, the Osmosis community has explored multiple strategies to mitigate MEV risks, including threshold encryption of pending transactions to prevent information leakage \cite{dev-yt-tenc-mempool,sunny-yt-threshold-enc-mempool,osmosis-zkfm,osmosis-rnd}. Additional techniques have also been discussed as complements, such as joint block proposals \cite[Slide 58]{dev-zksummit}, transaction order randomization, and batch clearing of trades to reduce price impact \cite{cow-protocol-batching}.

\medskip

These design choices involve tradeoffs that warrant continued analysis and debate within the community prior to implementation. This section aims to highlight some key considerations around the current MEV mitigation agenda on Osmosis, especially considerations related to the implementation of threshold encryption on the network.

\subsection{MEV Classification}\label{ssec:osmo-classification}

To the best of our knowledge, the MEV mitigation agenda is not yet finalized on Osmosis, but existing proposals give good insights as to how the Osmosis community plans to handle MEV on the network. According to~\cite{osmosis-7812}, the current approach to handle MEV on Osmosis consists of:

\begin{enumerate}
  \item Preventing 'harmful MEV', which is defined as: \textit{at minimum including actions that are based upon knowledge of ready-to-execute transaction content, prior to its execution}~\cite{osmosis-7812}. This includes behaviors like sandwiching, backrunning, censorship or delaying transactions, and
  \item Capturing 'benign MEV' (i.e.,~MEV that is not 'harmful') which refers to \textit{extraction based solely on the existing state or public information, such as arbitrage and liquidation}~\cite{osmosis-7812}. To do so, the Osmosis community proposes to capture 'benign MEV' via in-protocol mechanisms, to redistribute the proceeds to the community via a governance process.
\end{enumerate}

This direction seems reasonable conceptually. However, developing a robust methodology to categorize MEV toxicity remains challenging due to the many subtleties involved. Perspectives on harm can vary significantly between agents within the network ecosystem. As users can freely move cryptoassets across blockchains, overly strict limitations deemed detrimental by key participants like liquidity providers could drive an exodus of capital and activity to other networks more aligned with their profit motives. Thus, a balanced perspective accounting for diverse viewpoints will likely be needed, even if it means tolerating some MEV behaviors deemed harmful by certain constituencies.

One constructive path forward would be establishing a community-governed framework to quantify and rate the impact of various activities, using historical on-chain data to parameterize 'harm' based on impacts to user welfare, system overhead, liquidity effects, etc. (e.g., using an open methodology such as \cite{bloomberg-index-crypto}).

Constructing a robust toxicity rating methodology seen as fair and accurate remains challenging. However, such a data-driven approach could provide a more objective basis to distinguish acceptable from harmful MEV on an ongoing basis. We believe engaging the full spectrum of network participants, especially validators critical to security and operations, will be vital in devising solutions seen as legitimate.

\subsection{Slashing}

Current proposals suggest using slashing penalties to disincentivize validators from behaviors such as 'running harmful MEV extraction software', 'outsourced block production software', or 'custom transaction ordering software' \cite{osmosis-7812}. However, enforcing slashing for the full spectrum of possibly 'harmful' behaviors faces inherent difficulties:

\begin{itemize}
    \item Validators can disguise MEV activity across multiple accounts, or Sybils \cite{sybils}, to evade scrutiny. For instance, a validator may extract MEV via one account with no value at stake, while maintaining their validation role on another. Obscuring 'harmful MEV' activities using Sybils and benefiting from the information disclosure lag coming from the use of threshold encryption on the system could be used by malicious actors to, e.g., 
    make it difficult for other network members to gather enough data to find statistical patterns in decrypted blocks, in order to avoid or delay detection of malicious validator behavior.
    
    While social slashing \cite{social-slashing} (i.e., slashing by participant vote) may be possible in extreme cases, it involves a governance process, which may be hard to coordinate for frequent, harmful MEV-related activity that is hard to detect and quantify. Especially if not all voters have a technical financial background.
    \item Slashing is limited in its ability to disincentivize validators from MEV extraction. The potential profits likely exceed the cost of slashing penalties for a profitable attack. Slashing is bounded by the amount staked, whereas rewards for malicious strategies have no set upper bounds.
    \item Edge cases inevitably exist that are challenging to classify programmatically as permissible or malicious beyond doubt. Furthermore, the current definition of 'harmful MEV', defined as 'at minimum' can be seen as ambiguous and hard to enforce programmatically. For example, does threshold encrypted transaction spoofing in the Osmosis mempool qualify as 'harmful' under the current classification? The answer to this question looks unclear.
\end{itemize}

Moreover, activities that can be deemed harmful to the network may originate from any network user, not just validators. Non-validators have no staked assets at risk of slashing. For instance, a non-validator user may attempt to manipulate information related to Osmosis trading pools, by e.g., spoofing the mempool with 'dummy' encrypted transactions, or by voluntarily disclosing information about incoming trade intents (see, e.g., \ref{ssec:privacy-considerations}). Doing so with no 'value at stake' can allow influential traders to benefit from the market, without risking being penalized by the protocol\footnote{Repeated examples of market manipulation have been observed from firms (see~\ref{ssec:collusion}) or individuals, e.g.,~\cite{musk-manipulation}, which demonstrates the limits of 'reputation systems' to disincentivize such behavior}.

While such behavior may or may not be deemed 'harmful' under Osmosis' categorization framework (see~\ref{ssec:osmo-classification}), it is possible to foresee scenarios where behaviors that qualify as 'harmful' are possible, all while escaping slashing. Malicious wallet developers, for instance, could sandwich the trades of their users, allowing sandwiching activity to occur, even if the network uses threshold encryption. In this case, the wallet software would be able to access the transaction data before it gets sent on the network (i.e., before it reaches the mempool) or other information like IP addresses and the user interface used to generate the transaction \cite{walletSnitching}, and would allow non-validator wallet developers to carry out 'harmful' MEV while evading slashing penalties or causing misattribution of disallowed behaviour to certain validators.

\subsection{Governance}

The Osmosis improvement proposal \cite{osmosis-7812} mentions that Osmosis should implement in-protocol systems to mitigate harmful MEV while transparently redistributing profits from non-harmful MEV to the community. Assuming consensus on classifying benign MEV, a question is how such captured profits should be redistributed.

\medskip

The implemented \emph{ProtoRev} Module Community Proposal~\cite{osmosis-7078,osmosis-skip-protorev} added a module to Osmosis which automatically calculates and executes arbitrage after user AMM transactions, rebalancing Osmosis liquidity pools. To maintain technical efficiency, the module does not exhaustively search for all possible arbitrage, but aims to capture the majority of opportunities. The proceeds of this arbitrage will form a new type of protocol revenue and will then be redistributed to the community.

ProtoRev offers interesting tradeoffs, as it makes full use of the particularities of Osmosis' bounded MEV domain and application-specific focus on decentralized exchange
\footnote{Implementing a module like ProtoRev to comprehensively capture benign MEV opportunities for the diverse array of arbitrary smart contracts on generalized platforms like Ethereum presents a significantly more complex problem. This stems from the virtually infinite extraction space created by generic smart contract composability and Ethereum's stated goal of remaining neutral between deployed DApps.}.
This approach is pragmatic and creates an extra incentive to stake on the system (which improves the overall security of the network). Importantly, ProtoRev is compatible with the use of threshold encryption on Osmosis.

\medskip

However, reliance on governance for redistribution raises concerns around converting MEV problems into 'governance problems'. In \cite{osmosis-pob}, the authors mention \textit{``Governance has full control over block-building and payment distribution. The chain can vote how MEV is distributed among network participants, with payments enforced in consensus''}. Here, 'the chain' refers to its stakeholders.

In particular, influential stakeholders who are able to sway governance decisions may vote to redirect 'benign MEV' profits primarily to themselves rather than the broader community, all while colluding to decrypt the mempool and benefit from privileged information. This 'privatization' of on-chain MEV proceeds could exacerbate centralization forces that MEV mitigation aims to alleviate. The community should deeply consider solutions like ProtoRev that address MEV via governance. While pragmatic in the near-term, heavy reliance on redistribution voting risks recreating similar issues to the MEV challenges such approaches intend to resolve.

\subsection{Implementation of Threshold Encrypted Mempools}

\subsubsection{Design Considerations}

To the best of our knowledge, Osmosis' current plans to implement threshold encryption are best characterized by the Ferveo protocol~\cite{DBLP:journals/iacr/BebelO22}. Ferveo proposes efficient distributed key generation and threshold public key encryption schemes to encrypt and decrypt transactions on BFT consensus blockchains, aiming to achieve mempool privacy.

The protocol design addresses some, but not all, considerations raised in~\ref{ssec:privacy-considerations}:

\begin{itemize}
\item \emph{Message content \& DoS:} The protocol encrypts all transaction contents to all network participants until finalized in a block, except for public fees. This prevents spam by ensuring adversaries pay per transaction, and keeps blocks within gas limits. However, lacking full encryption reveals information and may enable selective censorship (see, \ref{ssec:privacy-considerations}), since public fees may indicate transaction intent. Fully encrypting transactions would likely create extensive ripple effects, necessitating significant protocol redesign. Given the current network state, these partial encryption tradeoffs seem inevitable.
\item \emph{Optionality:} Ferveo's design does not seem to enable optional disclosure of trade intent. Optionally disclosing (correct or misleading) trade information remains possible through revelation of plaintext corresponding to ciphertext, public unenforced claims about trade intents or the use of trusted third parties attesting to attributes of submitted encrypted transactions (as discussed in \ref{ssec:privacy-considerations}). However, users lack an avenue for making credible claims about trading intent without the use of centralized trusted parties. It remains unclear whether such out-of-protocol disclosure and actions resulting from this disclosure would qualify as a 'harmful' endeavor under Osmosis' MEV classification\footnote{By disclosing one's (genuine or misleading) trade intentions one can induce market reactions and thus information streams in the mempool. In such case, inferring information about the state of the mempool may be possible – despite threshold encryption -, which may allow extracting value. If the trade intention is honest, an influential market participant can get the rest of the network to coordinate to provide him with high levels of liquidity or best execution (e.g., a 'whale' announcing a large sell on an AMM may induce LPs to provision liquidity to absorb the trade and minimize price impact, effectively allowing the 'whale' to get the 'network to work for them'). If the trade intention is dishonest (i.e., 'fake news'), the malicious market participant can manipulate the market, for e.g., get one side of the market to buy (resp. sell) at poor prices.}. We refer readers to \ref{ssec:privacy-considerations} for a more detailed discussion. 
\end{itemize}

Beyond securely implementing encrypted mempools while preserving efficiency, using Ferveo may impact validator incentives. Osmosis' plans empower validators with more responsibilities (see, e.g., \cite{osmosis-github-validator}) like generating decryption keys and decrypting encrypted transactions~\cite[Section 4]{DBLP:journals/iacr/BebelO22}. This may lead validators to revise their commission structures to account for additional duties. Such changes could impact overall network incentives.

\subsubsection{Limitations}

\ref{sec:limits} outlines several areas in which the implementation of threshold encrypted mempools may fall short of desired goals or have unintended consequences. In order to better understand the implications of such a scheme for Osmosis, we turn to these areas with the specifics of Osmosis in mind. 
\begin{itemize}
    \item \emph{Collusion:} At present, just 7 validators, within the total pool of 150 active validators, cumulatively control over 34\% of voting influence \cite{mintscan-osmosis}. This high centralization raises concerns regarding the potential impacts of introducing threshold encryption, which invites and is susceptible to inconspicuous collusion.
    \item \emph{Ordering \& Information Asymmetry:} Osmosis' current design, with or without encrypted mempools, does not limit control over ordering. Similarly, the impact of information asymmetry and risk of decryption games as described in \ref{sec:limits} apply directly to Osmosis.
    \item \emph{Cryptographic Assumptions:} Ferveo relies on cryptographic assumptions that, if added to Osmosis, would expand the set of assumptions of the network, rather than fully align with existing assumptions. In fact, to the best of our knowledge, the current implementation of Osmosis~\cite{osmosis-github} does not make use of cryptographic pairings, while Ferveo does. Furthermore, we believe that the willingness of the authors to use weighted decryption shares to align with the Tendermint's security assumptions paves the way for other misalignment, and collusion.
    \item \emph{Censorship:} Although the concerns around broad censorship still apply to Osmosis, designs like \cite{multiplicity} are created with Tendermint chains in mind and promise improved censorship resistance, albeit still subject to collusion concerns. 
    \item \emph{Economic Inefficiency \& Additional Transactions:} The main idiosyncrasy of Osmosis with regard to efficient transaction ordering is ProtoRev. An analysis of the economic impact of protocol-enforced arbitrage falls outside the scope of this paper. However, some superficial statements are worth making. Firstly, ProtoRev's rebalancing reduces the sensitivity of pool prices to volume traded, potentially reducing the impact of ordering on execution quality. The intuition behind this is that without rebalancing, a transaction may trade against only one pool, moving that price much further than the optimal routing would have done.
    
    Secondly, due to ProtoRev's limitations, there is still incentive to speculatively submit arbitrage transactions. ProtoRev does not rebalance along all possible routes, nor does it take into account the price of assets on other domains like CEX's or other chains. As such, the concerns of blocks filled with failed transactions highlighted in \ref{ssec:additionalTx} merit further investigation.
    \item \emph{Network Size:} The proposed scheme should not impact the size of the validator set at the present moment, although the scheme might impact the feasibility of potential future changes, such as decreased block times or an expanded validator set.
    \item \emph{Risk Management and Market Impact:} As Mars Protocol requires outside actors to liquidate bad debt and Osmosis is clearly intended as a home for trading, the considerations in \ref{ssec:additionalCons} apply directly.
\end{itemize}

\subsection{Summary}

Based on the analysis carried in this paper, we conclude this section by providing some key suggestions as Osmosis considers implementing threshold encryption.

First, we encourage refining the classification of MEV on Osmosis. Defining 'harmful' MEV as 'a minimum' leaves ambiguity around behaviors not clearly 'harmful' or 'benign'. Clarifying these definitions would allow deciding on best measures to mitigate harmful MEV. It would also enable better assessing if/how threshold encryption aligns with the MEV mitigation agenda.

Second, we strongly encourage carefully evaluating collusion risks. The difficulty of detecting decryption collusion, coupled with the small number of influential validators on the network, warrants active discussion of threshold encryption's risks and benefits. This is especially important as Osmosis seeks to empower validators with more duties over time \cite{osmosis-github-validator}. Over-empowering validators may raise centralization pressures, incentivizing collusion for profits at later stages.

Third, when implementing threshold encryption, Byzantine scenarios around collusion, spoofing, etc. could be modeled on an incentivized testnet to empirically gather data. Teams of participants could voluntarily collude, manipulate information flows, and stress test the network in return for OSMO rewards. Likewise, validators could run specialized software to collect metrics, and malicious wallets could be implemented to model for some threats mentioned in this section. With client audits and testnet data, the community would be better equipped to decide if and how to deploy threshold encryption.

Finally, we encourage further exploration of modifications to the threshold encrypted mempool design that grant users greater control over the information which they choose to reveal.

In summary, implementing threshold encrypted mempools necessitates evaluating impacts on information flows, incentives, and harmful collusion potential. We believe this research can foster nuanced discussions within and beyond Osmosis, surfacing both drawbacks and advantages to enrich existing debates.

\end{document}